\documentclass[12pt,preprint]{aastex}

\newcommand{\co}{\mbox{\rm $^{12}$CO}}
\newcommand{\cothree}{{\rm $^{13}$CO}}

\newcommand{\cii}{\mbox{\rm [\ion{C}{2}]}}

\newcommand{\hi}{\mbox{\rm \ion{H}{1}}}
\newcommand{\hii}{\mbox{\rm \ion{H}{2}}}
\newcommand{\htwo}{\mbox{\rm H$_2$}}
\newcommand{\jone}{\mbox{($1\rightarrow0$)}}
\newcommand{\jtwo}{\mbox{($2\rightarrow1$)}}

\newcommand{\percmcu}{cm$^{-3}$}
\newcommand{\percmsq}{cm$^{-2}$}

\newcommand{\kmpers}{\mbox{km~s$^{-1}$}}
\newcommand{\Kkmpers}{\mbox{K~km~s$^{-1}$}}

\newcommand{\xcounits}{\mbox{cm$^{-2}$ (K km s$^{-1}$)$^{-1}$}}
\newcommand{\Ico}{\mbox{\rm I$_{\rm CO}$}}

\newcommand{\xco}{\mbox{\rm X$_{\rm CO}$}}
\newcommand{\aco}{\mbox{$\alpha_{\rm CO}$}}
\newcommand{\av}{\mbox{\rm A$_{\rm V}$}}

\newcommand{\vlsr}{\mbox{V$_{\rm LSR}$}}

\begin{document}

\title{Unusual CO Line Ratios and kinematics in the N83/N84 Region of
the Small Magellanic Cloud}

\author{Alberto D. Bolatto\altaffilmark{1,2}, Adam Leroy\altaffilmark{2},
Frank P. Israel\altaffilmark{3}}
\and
\author{James M. Jackson\altaffilmark{4}}

\altaffiltext{1}{bolatto@astro.berkeley.edu} 

\altaffiltext{2}{Radio Astronomy Laboratory, Department of Astronomy,
University of California at Berkeley, 601 Campbell Hall, Berkeley, CA
94720-3411}

\altaffiltext{3}{Sterrewacht Leiden, P.O. Box 9513, NL 2300-RA Leiden,
The Netherlands} 

\altaffiltext{4}{Institute for Astrophysical Research, Department of
Astronomy, Boston University, 725 Commonwealth Ave., Boston MA
02215}

\begin{abstract}
We present new CO \jone\ and \jtwo\ observations of the N83/N84
molecular cloud complex in the south--east Wing of the Small
Magellanic Cloud (SMC).  While the \jtwo/\jone\ integrated line
brightness ratio (in temperature units) is uniformly 0.9 throughout
most of the complex, we find two distinct regions with unusually high
ratios \jtwo/\jone$\gtrsim2$. These regions are associated with the
N84D nebula and with the inside of the 50~pc expanding molecular shell
N83. This shell is spatially coincident with the NGC~456 stellar
association and the HFPK2000--448 radio continuum/X--ray source
tentatively classified as a supernova remnant. We explore possible
causes for the high ratios observed and conclude that the CO emission
probably arises from an ensemble of small ($R\sim0.1$~pc), warm
($T_g\sim40$~K) clumps. Analysis of the CO shell parameters suggests
that it is wind--driven and has an age of slightly more than 2 million
years.  We have also used this dataset to determine the CO--to--\htwo\
conversion factor in the SMC, an especially interesting measurement
because of the low metallicity of this source ($\approx1/9$ solar).
Surprisingly, after comparing the CO luminosities of clouds in N83/N84 with
their virial masses, we find a CO--to--\htwo\ conversion factor $\xco$
only $1.9$ times larger than what we obtain when applying the same
algorithm to solar metallicity clouds in the Milky Way and M~33. This
result fits into the emerging pattern that CO observations with high
linear resolution suggest nearly Galactic values of \xco\ in a wide
range of environments.

\end{abstract}

\section{Introduction}
The Magellanic Clouds ---with their unobscured lines of sight, their
small internal extinction, and their profusion of star-forming
sites--- constitute excellent laboratories for the study of the
interaction between massive stars and their environment. They are also
the nearest metal--poor systems with active star--formation \citep[
$Z_{LMC}\sim Z_\odot/4$ and $Z_{SMC}\sim Z_\odot/9$ for the Large and
the Small Cloud respectively;][]{DU84,KD98}. As such, they provide
invaluable insight into the physics and chemistry of the interstellar
medium (ISM) in primitive galaxies.

In this work we will discuss some of the molecular gas properties and
kinematics in the N83/N84 region of the Small Magellanic Cloud
\citep[][ Figure \ref{optical}]{HE56}.  This region is especially
interesting because it is one of the few isolated, yet relatively
active star-forming regions in the otherwise inconspicuous SMC
Wing. It comprises several CO clouds \citep{MI01,IS03}, \hii\ regions
\citep{HE56,DEM76}, various early--type stars \citep{TL87}, and an
expanding shell coincident with both the NGC~456 stellar association
and a possible supernova remnant \citep[source 448 in][henceforth
\objectname{HFPK2000-448}]{HFPK00}. This expanding shell appears to
be interacting with the nearby molecular cloud, sweeping and
compressing an important mass of molecular hydrogen (at least $10^4$
M$_\odot$; c.f., \S\ref{virial}), and perhaps setting up a second
generation of star formation. If N~83 indeed contains a supernova
remnant, it represent one of a handful of opportunities to study the
interaction between a molecular cloud and a SNR in a metal poor
environment.

Recently, \citet{BO00} have reported the detection of unusually high
\co\ \jtwo/ \jone$\sim3$ ratios in the molecular envelope of the
N159/N160 complex in the Large Magellanic Cloud (LMC).  Such high
ratios are rarely observed in the Galaxy and are interesting because
they convey information on the structure of the ISM in these systems
(c.f., \S\ref{discrat}).  Bolatto et al. suggested that these high
ratios are a direct consequence of the low metallicity of the
medium. If this is the case, we expect them to be more pronounced in
regions with strong UV radiation fields and even lower metallicity.
We undertook the present study in order to search for anomalous \co\
\jtwo/\jone\ ratios in a region of very low metallicity, and thus to
verify how widespread such ratios are. A second motivation was to
attempt to further distinguish between the mechanisms that can cause
such ratios.

The present study also allows us to revisit the problem of estimating
molecular masses using CO observations in metal--poor systems. Low
heavy--element abundances cause lower dust--to--gas ratios in a low
metallicity ISM \citep[$D/G \sim Z^{1-2}$;][]{LF98}. Because dust
particles are efficient absorbers of UV light, they shield molecules
from dissociative radiation.  Thus, the lower dust--to--gas ratios in
the SMC \citep[$D/G_{SMC}\sim 1/30\,D/G_{MW}$;][and references
therein]{ST00} lead us to expect higher rates of molecular
photodissociation. This is particularly true for molecules such as CO,
where lower metallicities also imply lower column densities, hence
significantly smaller self-shielding. Indeed, it has long been known
that emission from molecules is much fainter in dwarf galaxies, which
tend to be low metallicity systems \citep[e.g.,][]{TA98}.

Because of these effects, exactly how CO emission traces the molecular
ISM in metal-poor systems remains uncertain, and it has stayed a goal
of extragalactic CO observations to characterize this
relationship. Since the early days of CO observations in the Galaxy
astronomers have derived CO--\htwo\ calibrations and used them to
compute \htwo\ column densities and masses \citep[e.g.,][and
references therein]{SSS84,BL86,DHT01}. The factor to convert \co\
integrated intensities into \htwo\ column densities is referred to as
$\xco={\rm N(\htwo)/I(\co)}$, and has been determined using the virial
theorem, X--ray shadowing experiments, and an array of other
techniques. As it is notoriously difficult to incorporate all of the
relevant physics (such as cloud structure and optical depth) into
model calculations, astronomers have relied on empirical calibrations
of \xco\ rather than on a--priori determinations.

Since its introduction, it has been clear that \xco\ must be a
function of the local conditions of the ISM or at least the average
properties of the cloud ensemble; some of the most relevant quantities
are thought to be radiation field, dust--to--gas ratio, and
metallicity.  There are several calibrations of \xco\ with metallicity
available in the literature
\citep{DH89,WI95,AST96,IS97,IS00,BJI99,BA00,BLG02}. These all agree
that decreases in ISM metallicity and dust--to--gas ratio should
result in noticeable increases in \xco, meaning a lower CO integrated
intensity per unit \htwo\ column density.  They differ, however, in
their estimates just how large these increases in \xco\ will be. In
particular, observational studies find that \xco\ appears to depend
strongly on the attained linear resolution.  Studies based on
observations with high linear resolution, such as that provided by
interferometers, tend to find smaller increases in \xco\ in
metal--poor galaxies than those accomplished with larger beams
\citep[e.g.][]{WI95,WA01,WA02,RO03}.  Based on comparisons with far
infrared dust emission, \citet{IS00} has ascribed these effects to
intrinsic observational bias.

\section{Observations}

\begin{figure}[p]
\plotone{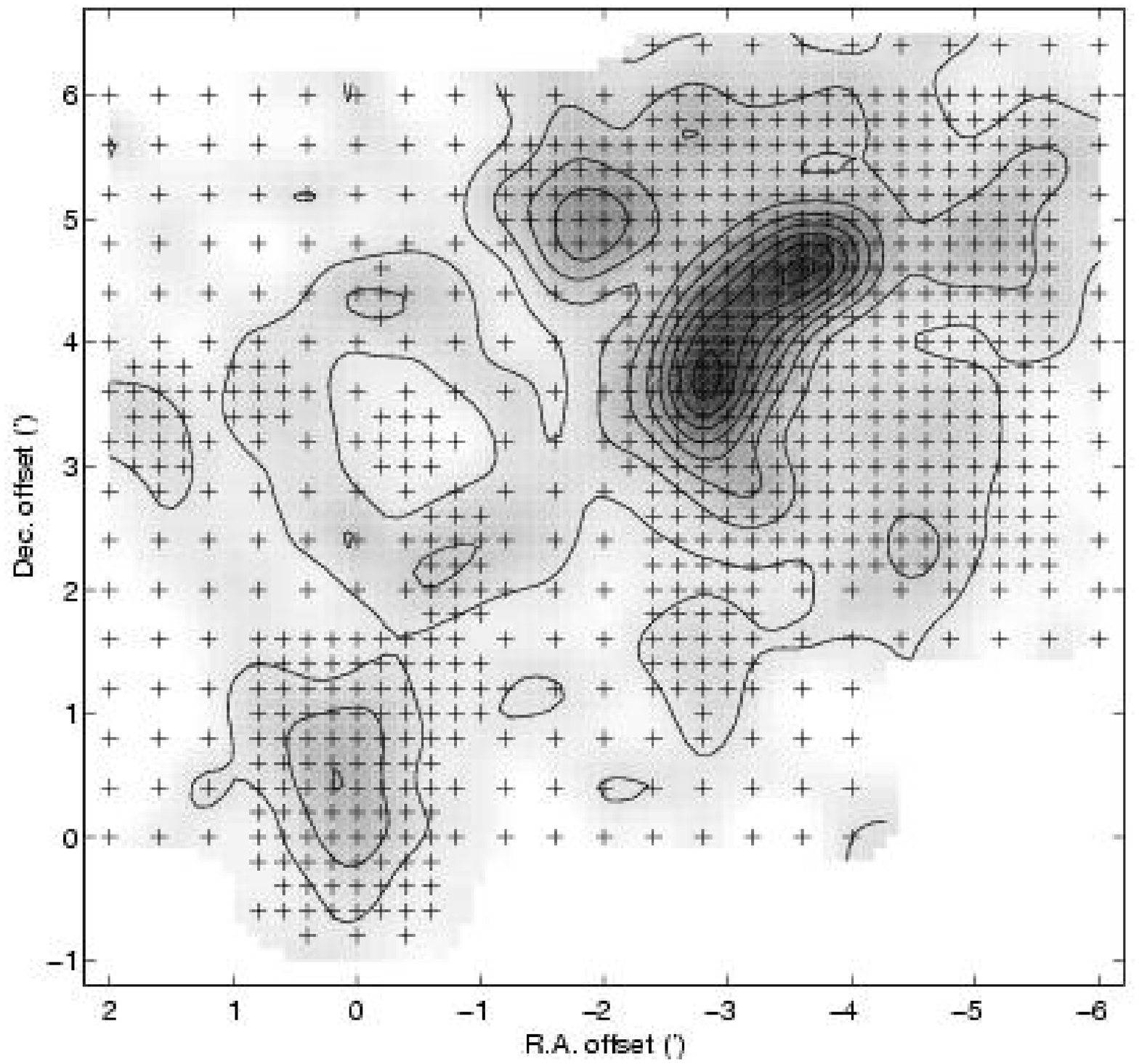}
\caption{Integrated intensity CO \jtwo\ map, showing the sampling. The
contour levels start at 1 and go in steps of 1 to 8 \Kkmpers. This map
was produced using gaussian interpolation, smoothing to a final beam
size of 38\arcsec\ HPBW. The offsets are with respect to
$\alpha_{1950}=1^{\rm h}13^{\rm m}23\fs1,
\delta_{1950}=-73^\circ36\arcmin33\arcsec$
($\alpha_{2000}=1^{\rm h}14^{\rm m}45\fs3,
\delta_{2000}=-73^\circ20\arcmin42\arcsec$).\label{mommap}}
\end{figure}

\begin{figure}[p]
\plotone{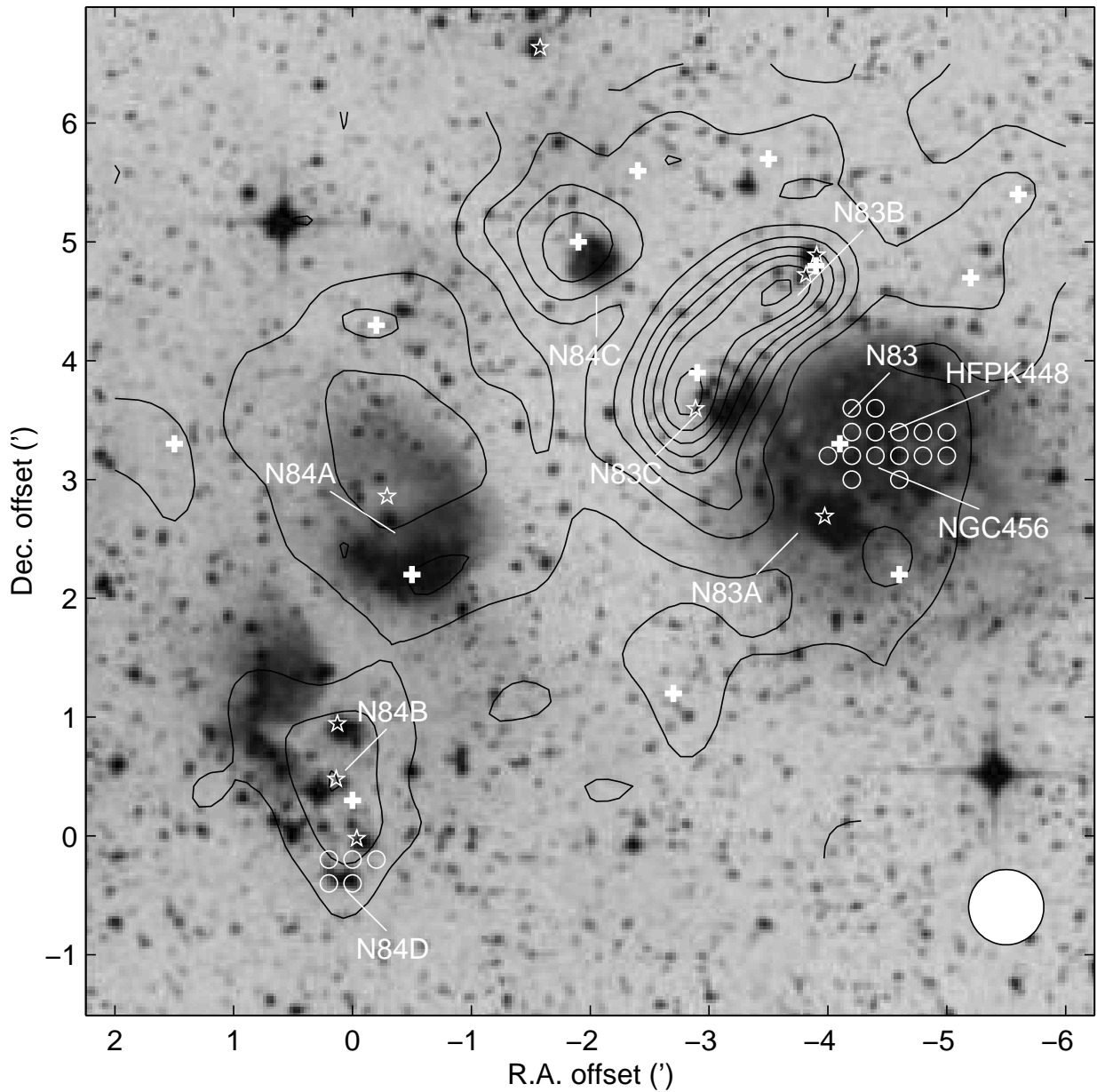}
\caption{CO \jtwo\ integrated intensity map overlaid on the annotated
DSS optical image of the N83/N84 region. The different symbols
indicate: IR sources in the SIMBAD database (stars), molecular clouds
listed in Table \protect\ref{cloudtab} (crosses), and individual
pointings with high \co\ \jtwo/\jone\ ratio (circles). The labels
indicate the approximate positions of the emission line nebulosities
listed in the Henize (1956) catalog, as well as two sources associated
with the expanding molecular shell: the NGC~456 stellar association
and the HFPK2000-448 radio continuum peak/SNR. The white circle in the
bottom right corner indicates the resolution of the CO map, identical
to that of Figure \protect\ref{mommap}.\label{optical}}
\end{figure}

Carbon monoxide emission from the N83/N84 region was first detected
and subsequently mapped by Israel et al. (1993, 2003) as part of the
ESO-SEST Key Programme ``CO in the Magellanic Clouds.'' Because of
considerable improvements in millimeter receiver sensitivity in the
years since, it has become possible to use SEST
\footnote{The Swedish-ESO Submillimetre Telescope (SEST) is operated
jointly by the European Southern Observatory (ESO) and the Swedish
Science Research Council (NFR).}  to make large maps of relatively
faint sources in a reasonable period of time. It has also become
possible to simultaneously observe the \jone\ and \jtwo\ transitions
of CO, allowing us to obtain reliable line ratios. We mapped a region
of $\sim10\arcmin\times9\arcmin$ encompassing N83, N84, and
HFPK2000-448, simultaneously at 115.271204 and 230.542408 GHz (Figure
\ref{mommap}). We detected over a dozen clouds in this area.  The
properties of the peak \jtwo\ spectra from these clouds are compiled
in (Table \ref{cloudtab}).

We acquired the maps in November 2000, by using SEST 115/230 GHz dual
SIS receiver and feeding the two frequencies into the split
high--resolution acousto--optical spectrometer (AOS) backend. The
resulting bandpasses ($\sim50$ \kmpers\ for 230 GHz and $\sim100$
\kmpers\ for 115 GHz) easily accomodated the narrow lines and small
velocity variations inside the molecular complex.  The spectra were
obtained using double--beam--switching mode, with a chop of
11.5\arcmin\ in azimuth. The few spectra contaminated by emission in
the off position were reobserved at a different hour angle. According
to its documentation, SEST has HPBW(230)$\approx23\arcsec$ and
HPBW(115)$\approx45\arcsec$, and its main beam efficiencies are
$\eta_{mb}(230)\approx0.50$ and $\eta_{mb}(115)\approx0.70$.  We
initially mapped a 24\arcsec\ grid (i.e., approximately full--beam
spacing at 230 GHz) and refined this to a 12\arcsec\ grid in regions
of bright emission and those regions in which we wished to increase
the signal--to--noise ratio. We used a nearby SiO maser to determine
pointing and focus corrections before each observing session. The SEST
pointing accuracy is 3\arcsec\ RMS. Assuming a 63 kpc distance to the
SMC, the scale of these observations is $\approx18$~pc per arcminute.

\begin{figure}[p]
\plotone{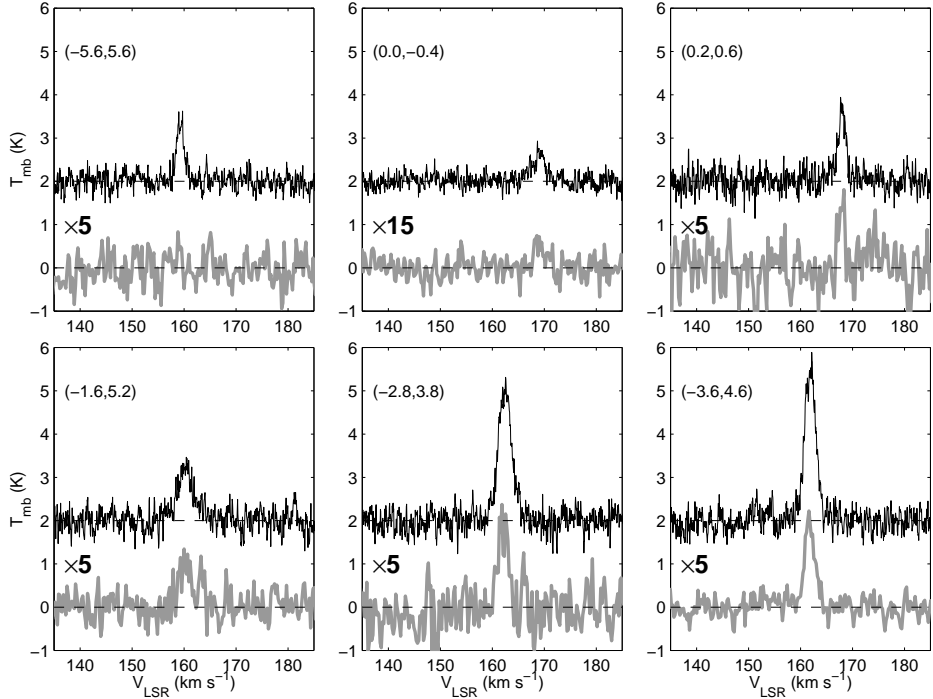}
\caption{\co\ and \cothree\ \jtwo\ spectra obtained near several peaks
in the integrated intensity map. The pointing near the N84D region
with offsets $(+0.0,-0.4)$, where we observe \co\ \jtwo/\jone\ ratio
$\sim2.0$, was also observed. This position was chosen because of its
relatively strong \co\ emission. Unfortunately, repeating the
\cothree\ observations toward other pointings with high ratios would
have required very long integrations.
\label{spectra}}
\end{figure}

To compare the \jtwo\ and \jone\ data, we convolved the spectra of
both transitions to a common angular resolution, thus removing the
effect of the different beam size of the observations.  We found
unusually high \co \jtwo/\jone\ line ratios in two regions of the map
(Figure \ref{optical}). The first instance occurs just south of N84,
close to the IRAS~01133--7336/N84D \hii\ region (Figure
\ref{hratspec}a). The second region is found in the gas inside the
expanding shell N~83, which is associated with NGC~456 and
HFPK2000-448 (Figure \ref{hratspec}b). To determine the optical depth
$\tau_{21}$ of the \jtwo\ emission we obtained \cothree\ \jtwo\
spectra toward several positions (Figure \ref{spectra}). Table
\ref{spectab} summarizes our observations, as well as our results for
$\tau_{21}$ obtained assuming N(\co)/N(\cothree)$\sim40-90$ (c.f.,
\S\ref{swc}).

\section{Discussion}

\subsection{The Origin of the Unusually High Line Ratios}
\label{discrat}

Optically thick CO emission arising from thermalized gas at $T>10$ K
yields a brightness-temperature ratio \jtwo/\jone$\sim1$.  Deviations
from these conditions ---for example, lower temperatures or subthermal
excitation in gas with volume densities below the critical density of
the \jtwo\ transition, $n_{crit}\sim10^4$ \percmcu--- generally
produce ratios {\it less} than unity. However, somewhat higher ratios,
($\sim1.2$) are frequently found in LMC and SMC molecular clouds
\citep[][and references therein]{IS03}. These are probably caused by a
mixture of high and low optical depths in molecular clouds and cloud
envelopes (i.e., a departure from the assumption of a homogeneous
source characterized by a single temperature and a single column
density).

Unusually high \co\ \jtwo/\jone\ ratios were observed by \citet{BO00}
in the LMC, who discussed four possible origins for them: 1)
self--absorbed \co\ \jone emission, 2) optically thin emission from
isothermal gas, 3) ensembles of small, optically thick isothermal
clumps, and 4) optically thick emission with temperature gradients.
With the available data they were not able to unequivocally determine
the cause of the high ratio, although self--absorbed CO emission
appeared unlikely. The same causes may explain the high ratios we find
in the N83/N84 region of the SMC. In this Section, we describe each of
these mechanisms and explore how well they can account for the new
data on the SMC.

\subsubsection{Self--absorbed emission}

Self--absorption is caused by an intervening envelope of colder CO in
the line of sight, which produces a dip in the background CO spectra.
This effect is more prominent for lower $J$ transitions, since most of
the intervening cold gas will be found near the ground state. The
spectra obtained toward the region south of N84 clearly show that the
\co\ \jone\ transition is not suppressed by self--absorption. Fig.~
\ref{hratspec} demonstrates that the two transitions have the same
line profile, thus convincingly showing they arise from the same
parcels of gas.

\subsubsection{Optically thin emission}

It can be shown that for gas in local thermodynamic equilibrium (LTE)
the ratio of optical depths in the \jone\ and \jtwo\ transitions is

\begin{equation}
\frac{\tau_{21}}{\tau_{10}} = 2 \frac{\left({1-e^{-\frac{2 h \nu_{10}}{k T_{ex}}
}}\right)}{\left({e^{\frac{h \nu_{10}}{k T_{ex}}}-1}\right)},
\end{equation}

\noindent where $\tau_{10}$ and $\tau_{21}$ are the opacities of the
respective transitions, $\nu_{10}$ is the frequency of the lower
transition, $T_{ex}$ is the excitation temperature, and $k$ and $h$
are Boltzmann's and Planck's constants. Since the opacity of
rotational transitions is proportional to the square of their rotation
quantum number $J$ at high temperatures, optically thin warm gas can
produce \jtwo/\jone\ ratios $\leq4$. The emission from such regions
will be faint, and observations in the Galactic plane may be biased
against finding them, which may explain why it is easier to find them
in the Magellanic Clouds. However, we clearly detected emission from
\cothree\ in the same place south of N~84 where we measured a high
\co\ \jtwo/\jone$=2.0\pm0.2$ ratio (Figure \ref{spectra}).  Although
the observed \co/\cothree\ \jtwo$\approx20$ ratio is somewhat high, it
nevertheless indicates that this transition is optically thick unless
the carbon isotopic abundance ratio is much lower in the SMC than in
the Galaxy. The best available determinations of isotopic abundance
ratios in the SMC suggest that N($^{13}$C)/N($^{12}$C)$\sim40-90$,
although data are scarce \citep{HE99}. In Table \ref{spectab} we
summarize our optical depth results for several clouds in the
complex. The range of $\tau_{21}$ values listed in the last column are
calculated assuming the quoted limits from \citet{HE99}.

For the pointing at ($+0.0,-0.4$) we find an opacity in the range
$\tau_{21}\sim1.5-4.4$, indicating that the \jtwo\ emission there is
optically thick. If the N(\co)/N(\cothree) column density ratio turns
out to be significantly lower than 40, however, the emission could be
optically thin. Such a scenario could conceivably result from the
combined effects of chemical fractionation and selective
photodissociation in the ISM. \citet{HE99}, for example, determined
this ratio to be $\sim15$ in the N27 region of the SMC using a mean
escape probability radiative transfer code modeling spherical clouds.

\subsubsection{Emission from very small warm clumps}
\label{swc}

\begin{figure}[p]
\plotone{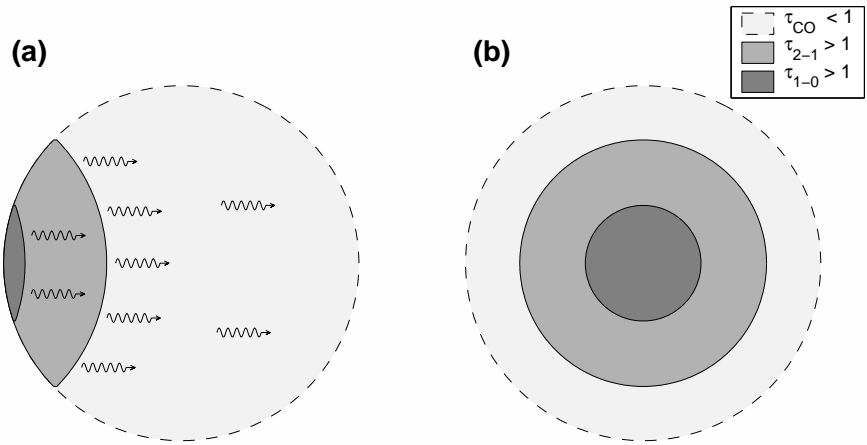}
\caption{A small, constant density, constant excitation spherical
clump model, showing the geometry of the $\tau=1$ surfaces for the CO
\jtwo\ and \jone\ transitions for an observer located (a) to the far right
of the clumplet, and (b) in front of the clumplet. The very
lightly shaded area surrounded by the dashed line is optically thin in
both transitions and contributes a small (albeit not negligible)
amount to the brightness of the clump in either transition. The darker
shaded areas indicate the regions where first the \jtwo\ and then
the \jone\ transitions accrue enough column density to become optically
thick. Most CO photons arise from these surfaces. Small, warm clumps
fill more of the beam in the higher, more opaque CO transitions.
\label{clumplet}}
\end{figure}

\begin{figure}[p]
\plotone{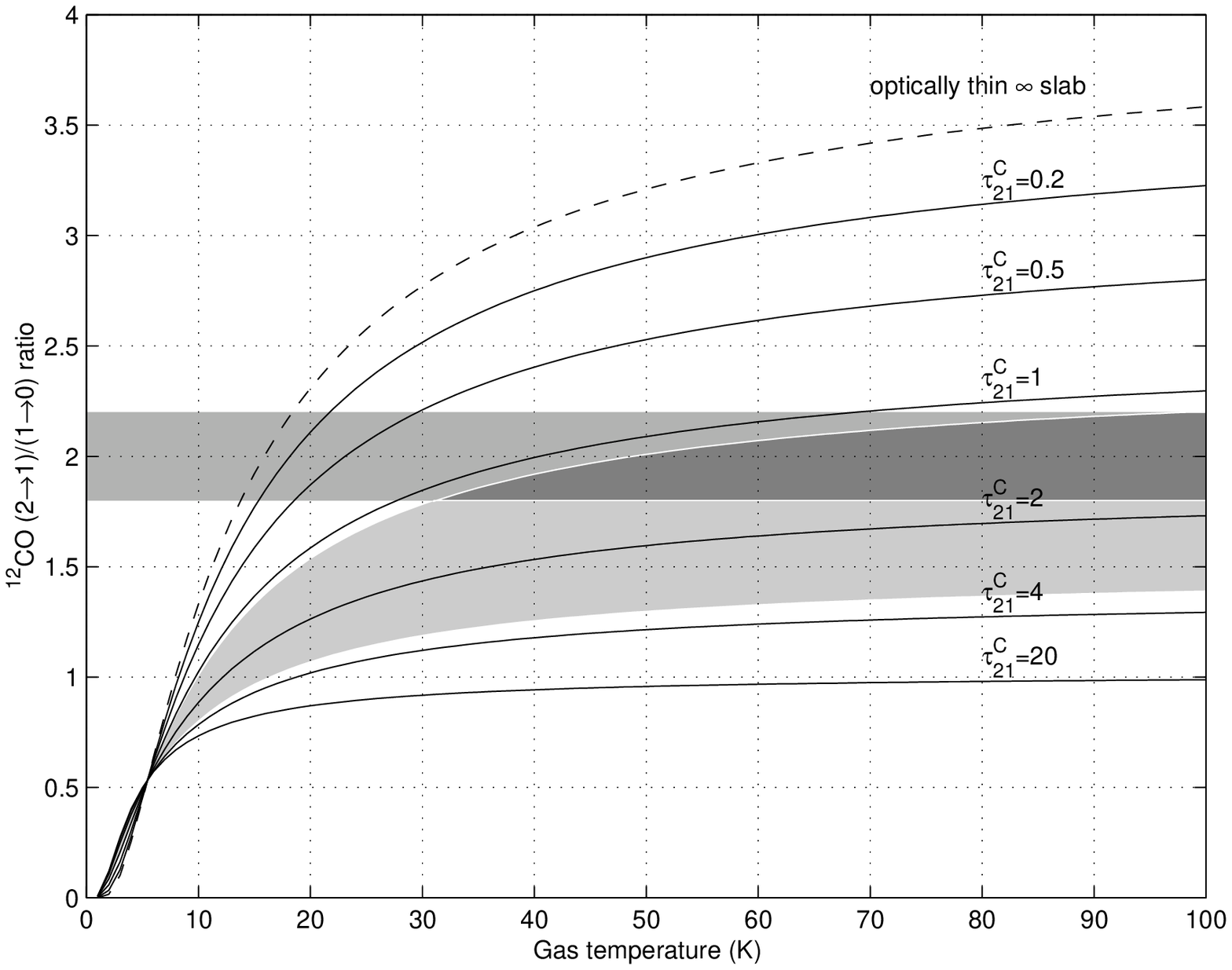}
\caption{CO \jtwo/\jone\ ratio for clumps of different central \jtwo\
opacity $\tau^C_{21}$, similar those of Figure \protect\ref{clumplet},
plotted against gas kinetic temperature. Note tha the overall opacity
of an spherial, constant density clump is 33\% higher than $\tau^C$,
which is the clump radius in opacity units.  The calculations use LTE,
assuming that the volume density is high enough to thermalize the
lower CO transitions ($n\gtrsim10^4$ \percmcu), and include the effect
of the optically thin portion of the clump. The shaded areas represent
the observational constraints for the region south of N83 where high
ratios were found, as discussed in the main text (\co\
\jtwo/\jone$\approx2$, $\tau_{21}\sim1.5-4.4$). \label{clumprat}}
\end{figure}

Collections of very small, isothermal, optically thick clumplets also
have the potential to produce high \co\ \jtwo/\jone\ ratios. They do
so partially through a beam filling effect; because of the faster
growth of the \jtwo\ opacity with column density the optically thick
region of the clumps appears larger in the higher transitions, thus
filling more of the beam. Figure \ref{clumplet} illustrates the
geometry of this effect, in which the optically thick $\tau_{21}=1$
``photosphere'' of the clump is larger and occurs closer to its
surface than the $\tau_{10}=1$ ``photosphere.'' Perhaps more
importantly, small clumps provide a natural geometry that allows the
\jtwo\ transition to become optically thicker than the \jone\
transition.  The line ratios for a spherical geometry can be easily
computed under the assumption of local thermodynamic equilibrium
(LTE); this is probably a safe assumption if the clumps have volume
densities in excess of $n\gtrsim10^4$ \percmcu\ (because of the
critical density $n_{crit}\propto J^3$, it is extremely difficult to
obtain a high \jtwo/\jone\ ratio with subthermal excitation unless the
equivalent radiation temperature is higher than the gas temperature,
which is not usually the case in mm--wave radio astronomy).  For each
transition, we assumed a constant gas temperature and density and
numerically integrated to find the opacity, $\tau(r)$, in a clump as a
function of $r$, the projected distance to the center. The results of
these calculations as a function of the gas temperature, $T_g$, and
the central clump opacity, $\tau^C$, are shown in Figure
\ref{clumprat}.  The shaded regions represent the observational
constraints: the measured \co\ \jtwo/\jone\ ratio and the central
clump opacity derived from the \co/\cothree\ \jtwo\ measurements as
described in the previous paragraph. The observations constrain the
gas temperature to be strictly $T_g>30$~K and more likely
$T_g\gtrsim40$~K. The clump radius is constrained to be $\tau^C\sim1$
in opacity units for the \jtwo\ transition. This opacity translates
into a physical column density $N(\co)\sim1.5\times10^{17}$ \percmsq\
at $T_g\sim40$~K, which can be converted to a length of $R\sim0.1$~pc
by assuming that all the carbon is locked in the CO molecules with a
carbon abundance of $\log({\rm C/H})\approx-4.7$ \citep{GA95,KD98} and
that the volume density is $n_{H_2}\approx10^4$ \percmcu\ (i.e., about
the critical density of the \jtwo\ transition). The resulting
parameters thus indicate fairly standard, but very warm, clumps.  The
heating is presumably provided by the star cluster found near the
high-ratio region \citep[source 166 of][]{BS95} or by the adjacent
\hii\ region and IR source (IRAS~01133--7336; Figure \ref{optical}).

The appeal of the ``small warm clumps'' scenario is twofold: 1) it
naturally explains why it is easier to find high line ratios in
systems of low metallicity, such as the LMC and SMC, and 2) as it has
predictive power, it can easily be falsified. Indeed, we have shown
that clump radii have to be small in opacity units to produce large
ratios. If the physical sizes of clumps do not change with
metallicity, low CO column densities will be more common in lower
metallicity systems because of the lower abundances of carbon and
oxygen. If the abundance of CO scales by a factor similar to that of
the metallicity, then a clump with central opacity $\tau^C(\co)\sim4$
in the Milky Way would have $\tau^C(\co)\sim0.5$ in the SMC. The
``small warm clumps'' model requires that the \cothree\ \jtwo/\jone\
ratio be $\sim3-4$ in this source, and it can be used to predict the
intensities of the higher CO transitions. The model also requires that
the N(\co)/N(\cothree) column density ratio be $\sim40-50$ south of
N84.

\subsubsection{Optically thick regions with large temperature gradients}

High brightness ratios can, in principle, also be produced by
optically thick gas if the $\tau=1$ surfaces of each transition occur
in regions of different temperature. To reproduce the observations, we
need a ratio of excitation temperatures
$T_{ex}(\tau_{21})/T_{ex}(\tau_{10})\gtrsim2$. How large a temperature
gradient does this change represent? Using the LTE approximation we
computed the column densities of CO necessary to attain $\tau=1$ for
each transition at different temperatures. We used this information to
obtain a very rough estimate of the typical increment in column
density $\Delta N(\co)$ required to go from the $\tau_{21}=1$ surface
at $T$, to the $\tau_{10}=1$ surface at $T/2$. We found that a
reasonable estimate is $\Delta N(\co)\sim 10^{16}$ \percmsq. Taking
into account the abundance of carbon, this is equivalent to a change
in visual extinction $\Delta \av\sim 0.05$. Since CO becomes the main
reservoir of carbon at $\av\gtrsim1$, this means that the cloud turns
optically thick in the \jone\ transition at a depth only $\sim5\%$
larger than that at which the $\tau_{21}=1$ surface is found. To
explain the observations, however, we require the excitation
temperature of the gas to drop by a factor of two for this $\sim5\%$
change in depth. Gas temperature profiles computed for PDRs externally
heated by UV sources show that such gradient is extremely unlikely
\citep[e.g.,][]{HT99}. This simple argument is also supported by line
ratios obtained from self--consistent PDR calculations.  At no point
in the parameter space of a homogeneous, externally heated PDR is the
CO $\jtwo/\jone>1.8$, and ratios greater than unity only occur at very
high densities and radiation fields \citep{KA99}. Perhaps embedded IR
sources with no UV flux, such as protostars, could produce the heating
necessary to obtain the right temperature ratio without the associated
photodissociation of CO.  This appears very unlikely, however, since
these sources would have to occur at a particular depth, coincident
with the $\tau_{21}=1$ surface.

We have explored four possible explanations for the unusually high
\co\ \jtwo/\jone\ ratios observed south of the N84 region. We have
shown that two mechanisms, self--absorption and optically thin
emission, are improbable. Of the remaining two, small warm clumps and
optically thick gas with large temperature gradients, we prefer the
first, although it is not possible to entirely discard the second with
the present data. High line ratios could also conceivably originate in
the optically thin portion of clumps with density gradients (for
example, critical Bonnor--Ebert spheres).  The proper modeling of such
objects, however, requires a full self--consistent radiative transfer
and temperature profile calculation, and is beyond the scope of this
paper: nevertheless we note that warm temperatures will still be
necessary to produce high line ratios.  In addition, the relevant
ratios are also obtained if we allow completely inhomogeneous
models. One of the simplest examples would be a population of cold,
dense clumps surrounded by a warm, tenuous envelope. Unfortunately, a
model containing only two completely independent components already
has a minimum of eight free parameters, which greatly exceeds the
number of constraining observational parameters (three in our case).
We may decrease the number of free parameters somewhat by making
plausible assumptions, but even in that case would we need more
constraints than we presently have. Future observations may remedy
this.

\subsection{The Expanding Molecular Shell}

\begin{figure}[p]
\epsscale{0.75}
\plotone{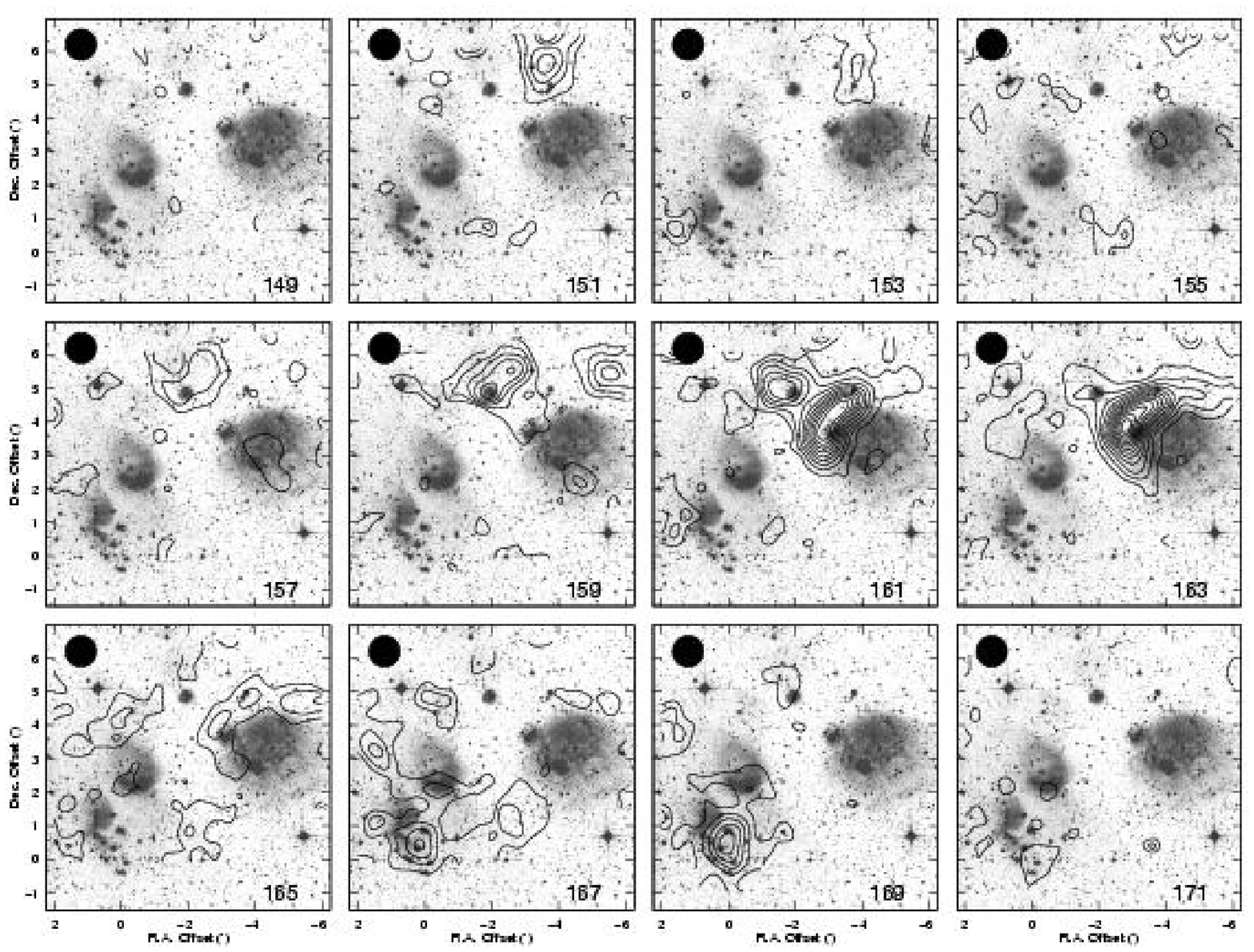} \\ \medskip
\plotone{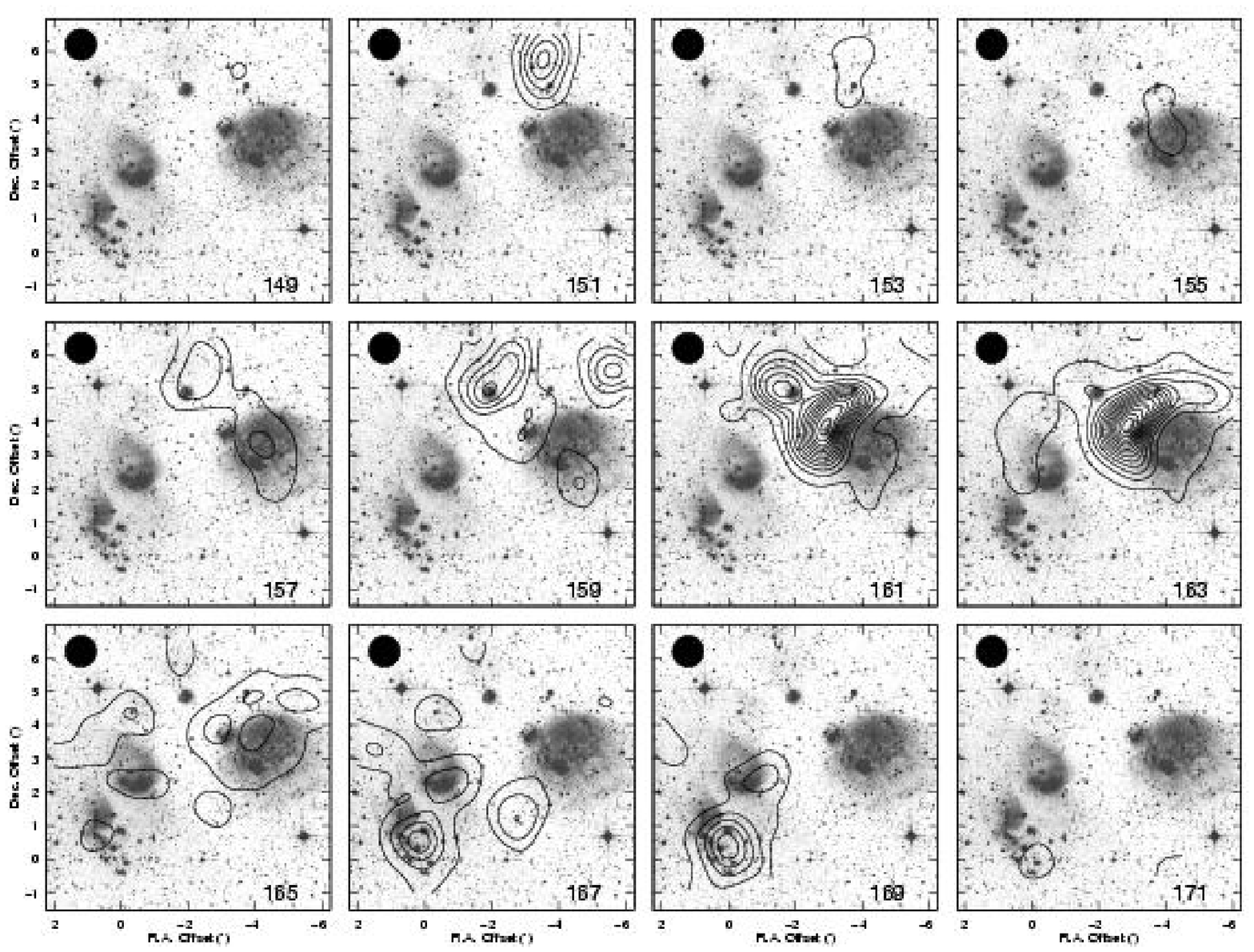}
\caption{Channel maps of the CO emission for the \jone\ (top group of
twelve panels) and \jtwo (bottom) transition, convolved to the same
angular resolution (55\arcsec) and overlaid on the DSS image of the
region. The contours are identical for both transitions: starting at
0.2 and increasing by 0.2 \Kkmpers. The ratio of main beam
temperatures is uniformly $\approx1$ throughout the map, except in the
two regions indentified in
Figure \protect\ref{optical}. \label{chanmaps}}
\end{figure}

The second instance of unusually high \co\ \jtwo/\jone\ ratios in the
N83/N84 region occurs in a remarkable place: the inside of the
expanding shell--like structure associated with NGC~456 and
HFPK2000-448 (Figure \ref{optical}). This optical shell has a
molecular counterpart apparent in a careful inspection of Figure
\ref{chanmaps}.  Starting at velocities 155--157 \kmpers\ the center
of N83 lights up, while at 161 \kmpers\ the eastern rim emits brightly
in CO. At 165 \kmpers, some emission shows up at the center of the
shell in the maps of the \jtwo\ transition. The spectrum obtained over
the central 1\arcmin\ of this structure even more clearly exhibits
this kinematic signature (Figure \ref{hratspec}b). While the emission
associated with the walls of the shell has very similar brightness
temperature in the \jone\ and \jtwo\ transitions, the gas {\it inside}
the shell appears brighter in the \jtwo\ transition by a factor of
$\sim2$.  The shell is $\sim50$~pc in diameter, and the molecular
material associated with it is expanding at a velocity $\sim\pm6.5$
\kmpers.  The northern and eastern edges of the shell are sharp and
well defined in the optical pictures, while its southwestern edge
appears ragged and soft. The cause of this difference in appearance is
strikingly clear in the CO channel maps: along the northern and
eastern edges, the expansion of the shell is contained by a molecular
cloud at $\vlsr\sim163$ \kmpers, while the south and west sides appear
able to expand more freely. It is not immediately clear whether this
bright CO cloud (corresponding to peaks MP~5 and MP~7 in Table 1)
consists of swept-up matter, or represents an ambient cloud
interacting with the expanding shell.

A question arises: is the expanding shell due to a supernova
explosion, or is it a wind--driven structure? The classification by
\cite{HFPK00} of HFPK2000-448 as a SNR was tentative, and based on
the detection of coincident radio continuum emission at 13~cm and
X--Ray emission by ROSAT. Extended X--ray emission and/or extended
nonthermal radio continuum are telltale evidence for SNRs. However, in
the case of HFPK2000-448 the data are not conclusive. The X--ray
emission could be extended but the quality of this determination (as
measured by its likelihood parameter) is low, and there is no
detection of the radio continuum at other frequencies to confirm that
the spectral index is nonthermal.

In any case, the rather modest expansion measured in CO is hard to
reconcile with the much higher expansion speeds expected from a
supernova remnant. It is quite in line, however, with the much lower
speeds that should characterize wind--driven shells. From the
interstellar bubble model by \citet{WE77}, we find that the age, $t$,
of the bubble is related to its radius, $R_{s}$, in parsecs and its
expansion velocity, $V_{s}$, in \kmpers\ by $t = 6 \times 10^{5}
R_{s}/V_{s}$. With a radius of about 25 pc and an expansion velocity
of 6.5 \kmpers, the age of the bubble should be about 2.3 million
years. From the model we also find the relation between mechanical
wind-luminosity, $L_{w}$, in erg s$^{-1}$ and original ambient total
hydrogen density, $n_{H}$, in \percmcu\ to be $L_{w}/n_{H} = 7.65
\times 10^{46} R^{5}/t^{3}$. By conservatively assuming $L_{w}/L_{bol}
= 10^{-3}$ for the ratio between mechanical wind and stellar
bolometric luminosities, we may reduce this to $L_{bol}/n_{H} = 6
\times 10^{37}\,{\rm erg s}^{-1} {\rm cm}^{3} = 1.5 \times 10^{4} {\rm
L}_{\odot} {\rm cm}^{3}$.  If the total stellar luminosity of the
embedded association NGC~456 is of the order of $10^{7}$ L$_{\odot}$,
the original ambient density should have been of the order $n_{H}
\approx 650$ \percmcu.  These luminosities and original densities seem
quite reasonable for a luminous association dispersing its molecular
birthcloud.  It is also not unreasonable to suppose that a massive
member of the association turned supernova within a few million
years. If such an event indeed occurred, the resulting remnant should
rapidly have filled the wind--blown bubble that we have observed.
Consequently, the chemistry and excitation around N~83 could be
dominated by SNR--caused shocks, even though the overall kinematics of
the shell would still be those of a wind--driven structure. We found
no evidence, however, of uncommon mm--line ratios in the N83B/C
molecular cloud. Its \co\ \jtwo/\jone\ ratio is $\sim0.9$, identical
to that of the rest of the complex, and our observations of HCO$^+$
\jone\ toward offsets ($-3.6,4.6$) yield a ratio \co/HCO$^+$
$\approx18\pm2$, similar to that reported by \citet{HE99} toward the
SMC source N27.

It would be of considerable interest to confirm the SNR identification
of N~83/HFPK2000-448. If it is confirmed, N~83 in the SMC would be
very similar to the well-established SNR N~49 in the LMC, which also
appears to be interacting directly with a molecular cloud
\citep{BA97}. N~83 would represent another excellent opportunity to
study molecular cloud--SNR interactions in a metal-poor environment,
complementing studies of similar interactions within the
Galaxy. Indeed, notwithstanding the fact that most Type II supernovae
must occur within their parent clouds and despite their potential
importance as triggers of a secondary wave of star--formation activity
\citep{OP53}, there are only a handful of such molecular cloud--SNR
associations identified in the literature \citep[e.g.,][and references
therein]{CCK77,WO77,WRM98,KI98,DU99}. Most of these interactions lie
in the Galactic plane, where obscuration and confusion pose
observational difficulties that are largely absent in the Magellanic
cluds.

\subsection{The \xco\ Factor in N83/N84}

\subsubsection{Measuring Cloud Properties}

To measure the individual cloud properties we proceeded as follows:
from the individual map spectra we constructed uniformly sampled main
beam intensity data cubes in the CO \jtwo\ and \jone\ transitions
using the COMB built--in 2D gaussian interpolation. The cubes have
angular resolutions (HPBW) of $35\arcsec$ and $55\arcsec$,
respectively, corresponding to spatial resolutions of $11$~pc and
$17$~pc. Their velocity resolution is 0.25 \kmpers. After convolving
the \jtwo\ cube to the resolution of the \jone\ cube, we carried out a
pixel--by--pixel comparison of the two cubes and found that the best
linear relation between both datasets was $\Ico\jone =
0.9\,\Ico\jtwo$, very similar to other results obtained for Magellanic
Cloud objects \citep[c.f.][and references therein]{IS03}. This scaling
will be used later to compute molecular masses using CO \jtwo\
observations.

We proceeded to identify clouds in the channel maps, and then measured
three properties for each cloud: its size, its velocity dispersion,
and its integrated CO intensity. From the first two quantities we
calculated a virial mass for each cloud, which we compared to the mass
derived from the integrated intensity using the Galactic CO--to--H$_2$
conversion factor. To obtain these numbers we specified a threshold
intensity, then built a mask that started from a seed position and
velocity, and included neighboring regions of the data cube with two
adjacent velocity channels above this threshold. We defined a
``cloud'' as an unmasked region contiguous in $\alpha-\delta-v$
space. This is a practical definition that is similar or identical to
those employed in a number of other studies
\citep[e.g.][]{SO87,HCS01,RO03}. The threshold intensity was varied
over a wide range, and cloud properties were computed for each of its
values. Obviously, very high values of the threshold will select only
small regions within a cloud, while low values will tend to merge
clouds together, and even lower values will include positive noise
as emission. We determined the reasonable range of values to use for
each cloud by observing when the cloud properties experienced a ``jump''
caused by the mask growing into a neighboring cloud.

After identifying an individual cloud, we calculate its size, velocity
dispersion, and CO flux as follows. To obtain the size, we compute the
second moment of the integrated brightness distribution of the cloud
in the $\alpha-\delta$ plane (i.e., the RMS size). Because the moment
underestimates the true size for gaussian clouds, we include a
correction factor that depends on the ratio of the threshold intensity
used to the peak intensity of the cloud. The inclusion of this factor
is important for two reasons: 1) in cases of confusion and low
signal--to--noise, it allows us to estimate the properties of an
entire cloud from the portion that is easily identifiable, and 2) it
avoids oversubtraction in our beam deconvolution. It assumes, however,
that the cloud is gaussian in position--velocity space. Fortunately,
this assumption can be veryfied by observing how the cloud properties
change as a function of threshold intensity (they should be constant),
and it appears to hold for the clouds here studied.  From the
corrected second--moment values, we ``deconvolve'' the beam by
subtracting its size from the measured cloud size in
quadrature. Finally, we apply the factor of $\frac{3.4}{\sqrt{\pi}}$
\citep{SO87} to convert from the RMS size to the radius of a
theoretical spherical cloud (this factor can be readily motivated by
considering a constant density spherical cloud and comparing the RMS
size to the spherical radius). Along with an assumed SMC distance of
63~kpc (see Walker 1998; Cioni et al. 2000), this treatment results in
a physical cloud size given by the formula:

\begin{equation}
R_{sph} \approx 0.41 \sqrt{\sigma_\alpha^2 + \sigma_\delta^2 - 2
\sigma_{beam}^2},
\end{equation}

\noindent where $\sigma_\alpha$ and $\sigma_\delta$ are the
threshold--corrected RMS sizes of the cloud in the $\alpha$ and
$\delta$ directions, $\sigma_{beam}$ is the RMS size of the beam (all
measured in arcseconds), and $R_{sph}$ is the spherical radius of the
cloud in parsecs.

Other works, in particular those focused on Milky Way clouds, have
treated clouds as ellipsoids and used the corresponding form of the
virial theorem due to \citet{BM92}. This approach is problematic in
extragalactic astronomy, because the observations do not have many
resolution elements across the cloud and their signal--to--noise is
low.  Under these conditions deconvolution becomes very difficult and
the precise shape of the source is hard, if not impossible, to
recover. Therefore, we content ourselves with measuring a single size
parameter and considering the simple model of a spherical cloud.

We calculate both the threshold--corrected RMS velocity dispersion,
$\sigma_v$, and the equivalent width, $EW$, from the integrated
spectrum of each cloud. For the equivalent width, we employ the
definition of Heyer et al.  (2001),

\begin{equation}
EW = \frac{\Sigma_i \psi(v_i)}{\mbox{Max} (\psi (v_i))}
\end{equation}

where $\psi (v_i)$ is the spectrum of the cloud in question. For a
gaussian distribution, $EW = 1.06\,FWHM$. Because it does not diverge
when signal--free noise is added to the edge of a spectrum, this
linewidth measure is more robust than the second moment as a measure
of the RMS velocity dispersion. Moreover, by comparing these two
measures we get a good idea of how well a spectrum is approximated by
a gaussian (for our data, the ratio $EW/\sigma_v$ correlates very well
with the $\chi^2$ value of a gaussian fit to a spectrum).

We compute the CO luminosity by summing the intensities of the entire
cloud and again applying a correction factor based on the
peak--to--threshold ratio and an assumed gaussian profile. In the case
of the CO \jtwo\ data cube, we applied the aforementioned $0.9$
conversion factor to calculate the corresponding CO \jone\ luminosity
for the cloud: essentially, we used the \jtwo\ measurements as
predictors of the \jone\ intensities, taking advantage of their higher
S/N and spatial resolution. To compute molecular cloud masses
(including the contribution by He) we use the formula

\begin{equation}
{\rm M_{mol}}=\aco\, S_{\rm CO} \,d^2,
\end{equation}

\noindent where $S_{\rm CO}=\int\Ico\,d\Omega$ is the spatially
integrated \co\ flux of the molecular cloud measured in Jy~\kmpers,
$\aco=1.07\times10^4$ $M_\odot$~Mpc$^{-2}$~(Jy~\kmpers)$^{-1}$
(corresponding to a canonical conversion factor $\xco = 2 \times
10^{20}$ \xcounits\ ), and $d$ is the distance to the SMC in Mpc.

We compute the virial masses using the assumption that each cloud is
spherical and virialized with a density profile of the form $\rho
\propto r^{-1}$. Thus the virial mass is given by the formula
\citep{SO87}

\begin{equation}
{\rm M_{vir}} = 1040\, \sigma_v^2 \, R_{sph},
\end{equation}

\noindent where $\sigma_v$ is the RMS cloud velocity dispersion in
\kmpers, $R_{sph}$ is the spherical cloud radius in pc, and M$_{\rm
vir}$ is the cloud's virial mass in M$_\odot$.

\subsubsection{Results}
\label{virial}

We have applied the procedure described above to several subsets of
the data, as well as the whole datacube. We focused our attention on
three clouds contained in the Nyquist--sampled regions of our original
dataset, roughly corresponding to MP7, MP10, and MP13 in Table
\ref{cloudtab} (i.e., the clouds associated with N83B/C, N84C, and
N84B/D respectively; Fig. \ref{optical}). We also computed masses for
what we call the ``northern region'' (the complex encompassing N83B/C
and N84C but not including MP6, which appears to be an unrelated cloud
peaking a much lower velocities), and for the complex as a whole. Our
results are summarized in Table \ref{cotab}.  Because the dominant
sources of uncertainty in our determination of cloud masses are the
cloud boundary definitions, we repeated the procedure described above
for a wide variety of thresholds. Thus, we probed the properties of
the complex on a variety of size scales, and on each scale we
estimated a reasonable range of values for each of the physical
parameters. In Table \ref{cotab} we give the properties of structures
on scales ranging from what seem to be compact, self--gravitating
object to complexes composed of a few large, bound, virialized clouds
connected by low surface brightness filaments.

How do the results obtained from the \jtwo\ data compare with those
from the more commonly observed \jone\ transition? CO \jone\ usually
provides the basis for estimating $\xco$ in the SMC and other
galaxies, even though the lower spatial resolution at $115$ GHz makes
size measurements more difficult. We performed identical analyses on
both datasets. For the most part, the $115$ GHz properties were
similar to those derived from the $230$ GHz data, though the CO flux
and linewidth of MP10 are both notably larger at $115$ GHz. The
complex, as a whole, has somewhat higher linewidth (and consequently
virial mass) when measured in the \jone\ transition. This is
artificial and caused by the higher noise of the \jone\ datacube: to
encompass the same area used in the \jtwo\ calculations we need to use
too low a threshold, causing positive noise to be included as
emission.

Because of the higher resolution and better signal--to--noise, we
prefer the results from the \jtwo\ data. Assuming all clouds share the
same conversion factor, these results yield $\xco\approx7.0 \pm 3.4
\times 10^{20}$ \xcounits\ (i.e., $3.5 \pm 1.7$ times the Galactic
value) on the size scale of individual molecular clouds ($\sim10$ pc).
Using this conversion factor, the total molecular mass traced by CO in
the complex is $M(\htwo)\sim 2.5 \times10^5$~M$_\odot$.

We believe our algorithm to be both robust and simple. It is designed
to be easily applicable to a heterogeneous set of
observations. However, because several details (particularly the
masking correction) differ from those previous works, our results
(particularly our size measurements) may not be readily comparable to
those in the literature. To account for this, we applied our algorithm
to both the Rosette Molecular Cloud \citep{WBS95} and a subset of well
resolved solar metallicity clouds from the BIMA survey of M~33
\citep{RO03}. Both of these tests yielded values for \xco\ of $3.4-4.0
\times 10^{20}$ \xcounits. In light of this, our result can be
restated as follows: our measurments of the \xco\ factor in the SMC
yield a value for \xco\ $1.9 \pm 0.9$ times the value of \xco\
obtained for identical measurements on the Rosette Molecular Cloud and
several M~33 clouds.

How do these molecular masses and estimates of \xco\ compare to those
obtained by other studies of the SMC? The results presented here are
consistent with those obtained independently by Israel et al. (2003),
taking into consideration that they mapped only part of the complex in
the \jone\ CO transition, with generally lower signal--to--noise
ratios. Our \jtwo\ result is lower by about a factor of $\sim 2$ than
the average value \xco$\sim12\times10^{20}$ \xcounits\ obtained by
\citet{RU93} from SEST CO \jone\ measurements of clouds elsewhere, in
the southwest Bar of the SMC. However, when we adjust the sizes of
\citet{RU93} to match our own definition and include beam
deconvolution, we find a $\sim 40\%$ reduction in the average size of
a cloud. The corresponding drop in cloud virial masses suggests a
value of \xco$\sim$ 6---8$\times10^{20}$ \xcounits\, consistent with
our estimates. We also take the opposite approach, bringing our
results into line with their measurements by considering the
properties of the N83 complex at the $0.4$ K \kmpers integrated
intensity contour. If we reject the low velocity emission, as
\citet{RU93} did for several clouds, and employ their definitions and
formulae then we obtain a conversion factor of \xco$\sim 20
\times10^{20}$ \xcounits, in excellent agreement with their average
result for complexes ($23 \times 10^{20}$ \xcounits).  This value is
also in agreement with that obtained by \citet{MI01} using the NANTEN
telescope, with a beam size HPBW$\approx2\farcm6$.

\subsubsection{Caveats and Limitations}

We note that most common limitations inherent to the data such as low
signal--to--noise, blended line profiles, and low spatial resolution
will drive our estimate of \xco\ toward higher rather than lower
numbers. All the usual caveats regarding the applicability of the
virial theorem apply, with the added concern that we are considering a
particularly active region. The effects of an expanding shell and
active star formation on nearby clouds, although not immediately
apparent from the data, may be important.

We should caution that virial estimates of \xco\ on the large scales
are suspect. First, they require assuming that the entire complex is
virialized and that large scale motions (e.g., galactic rotation,
expanding shells) are negligible. Second, the mass on the spatial
scale of the measurement has to be dominated by molecular hydrogen,
meaning that the contributions from stars and \hi\ are negligible. The
ATCA+Parkes HI observations of this region \citep{ST99}, for example,
show that the atomic hydrogen mass in a 120 pc diameter region
centered on the N83/N84 complex is $\sim530\times10^{3}$ M$_\odot$,
very similar to our virial mass estimate. Because this is an integral
along the line--of--sight, however, it is impossible to assert how
much of this mass is really associated with the complex. Third, they
require assuming that the velocity field inside the region is
uniformly traced by CO emission, otherwise the intensity--weighted
second moment usually employed to obtain the velocity dispersion is
meaningless.

Finally, we have gauged the magnitude of the conversion factor
indepently from virial considerations by following the method outlined
by \citet{IS97}. We use his SMC calibration $N_{\rm H}/\sigma(FIR) =
1.65\times10^{28} {\rm cm}^{-2} ({\rm W m}^{-2} {\rm sr}^{-1})^{-1}$
to derive the molecular mass from the far infrared emission near the
N83/N84 complex. At 15\arcmin\ resolution, the temperature--corrected
far--infrared surface brightness is $\sigma(FIR) = 7\times10^{-7} {\rm
W m}^{-2} {\rm sr}^{-1}$ \citep{SCH88,SI90} so that $N_{\rm H} =
1.2\times10^{22} {\rm cm}^{-2}$.  Both the old \hi\ map by
\citet{MN81} and the newer ATCA+Parkes data by \citet{ST99} yield a
mean neutral hydrogen column density of $N_{\rm HI} = 5.5\times10^{21}
{\rm cm}^{-2}$. From this we deduce the molecular hydrogen column
density $N_{\rm H2} = 3.2\times10^{21} {\rm cm}^{-2}$, corresponding
to a molecular gas fraction $2N_{\rm H2}/(N_{\rm H}+N_{\rm He}) =
0.4$, which is not unusual for a star--forming region. The CO
luminosity integrated over the N~83/N~84 complex is given by
\citet{MI01} as $L_{\rm CO} = 1.3\times10^{4} \Kkmpers {\rm
pc}^{2}$. As this luminosity is beam-independent for unresolved
sources, we can easily extrapolate to the 15\arcmin\ resolution of the
molecular hydrogen column density just derived. At that resolution,
the CO integral corresponding to the observed CO luminosity would be
$I_{\rm CO} = 0.22 \Kkmpers$.  This is probably an underestimate,
because such a large beam would encompass additional \hii\ regions
most likely associated with their own CO clouds. Our best estimate --
to be verified by future CO mapping of the area -- is therefore
$I_{\rm CO} = 0.33\pm0.11 \Kkmpers$, yielding an independent estimate
$\xco=100\pm50\times10^{20}$ \xcounits\, very similar to the other
\xco\ values of the SMC found in the same way \citep{IS97}.

These results fit well into an apparently consistent pattern.
Observations with high linear resolutions, i.e. single--dish
observations of nearby extragalactic objects such as those presented
here or interferometric observations of more distant objects,
consistently suggest virial cloud masses implying nearly Galactic
\xco\ conversion factors regardless of environment. By contrast,
observations not resolving individual clouds but covering whole
complexes have suggested much higher conversion factors in metal--poor
systems. This pattern may be explained in several ways: 1) it may be
caused by a large ``mm--dark'' molecular mass in metal--poor galaxies,
comprised by \htwo\ clouds without CO emission.  In this scenario the
observed CO emission arises only from the densest, most shielded
clouds in a complex. If this is true, CO observations suffer from a
severe observational bias.  2) It is possible that the complexes
observed are not virialized, or their masses dominated by
unaccounted--for stellar or atomic gas components, and that there is
not a great deal of hidden molecular mass. In this case,
high--resolution observation such as those presented here, suggest
that individual clouds are virialized with an approximately Galactic
\xco, indicating that metal--poor galaxies have molecule--poor ISMs.

If the complexes are virialized and their masses are dominated by the
molecular gas, then their global dynamics would be better probes of
the overall molecular mass than the dynamics of individual clouds. On
the other hand, if complexes are not virialized or their masses not
dominated by molecular gas, then there may be no way for CO
observations to get a handle on the total molecular masses of
metal--poor galaxies. In this case other tracers, such as
far--infrared emission from dust, large surveys of UV \htwo\ line
absorption, or FIR \cii\ measurements, may be the only way of
obtaining global molecular mass estimates.  By themselves, CO
observations are insufficient to distinguish between these two
scenarios. Observations of other tracers, such as the far--infrared,
will be necessary to establish whether CO is a good tracer in
metal--poor galaxies.

\section{Summary and Conclusions}

We have presented new CO \jone\ and \jtwo\ observations of the N83/N84
nebulae in the SMC that show an active and complex region, home to
several interesting astrophysical phenomena. We have identified 14
molecular clouds and an expanding molecular shell with a center
coincident with the NGC~456 stellar association and the HFPK2000--448
radio continuum/X--ray source classified by \citet{HFPK00} as a
SNR. The shell is most likely wind--driven, and appears to have an age
of 2.3 million years. We have also identified two extended regions of
unusually high \jtwo/\jone\ ratio: the first found south of the
complex near the N84D emission line region, the second located inside
the expanding molecular shell. A \cothree\ spectrum toward this first
region shows that the \jtwo\ transition has opacity $\tau\approx
1.5-4.4$ assuming N(\co)/N(\cothree)$\sim40-90$.

\begin{figure}[p]
\plotone{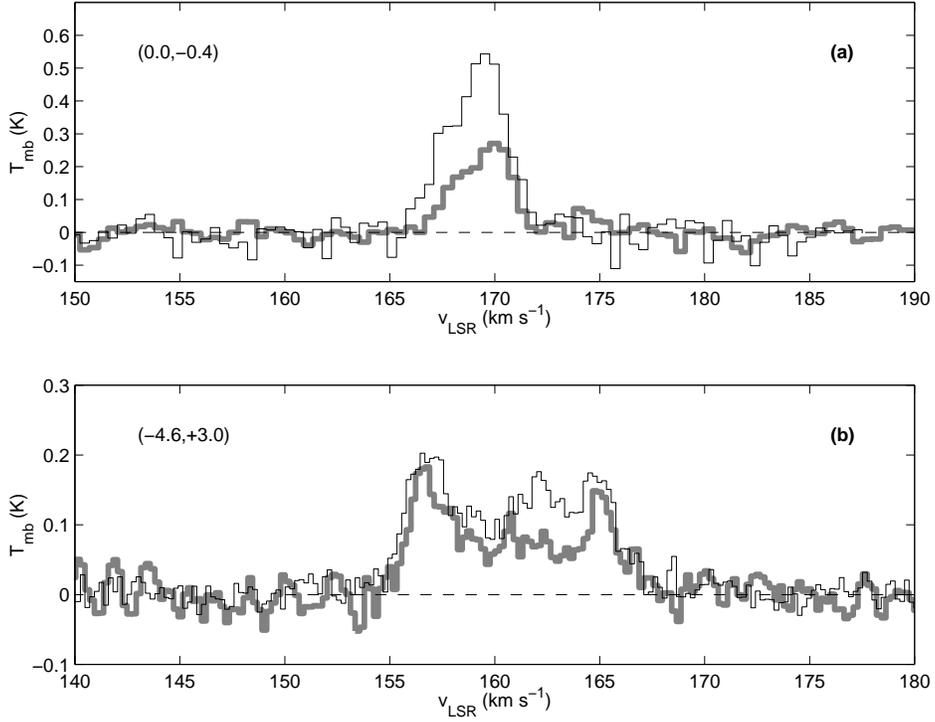}
\caption{{\em (a)} CO \jone\ (thick line) and \jtwo\ (thin line)
toward the region south of N84 (N84D), corrected by their respective main
beam efficiencies and with the higher transition convolved to the
angular resolution of the lower one. Remarkably similar velocity
structure, and a line ratio $\sim 2.0$. {\em (b)} Same toward the CO
expanding shell, with both transitions convolved to 1\arcmin\
resolution. The \jtwo\ transition is enhanced in the center component,
corresponding to gas filling the inside of the shell.\label{hratspec}}
\end{figure}

Barring the unlikely cases that the isotopic N($^{12}$C)/N($^{13}$C)
carbon ratio is considerably lower than in the Galaxy or that
fractionation effects have decreased the isotopomer N(\co)/N(\cothree)
ratio to much below the isotopic ratio, this high line ratio is not
caused by low optical depths.  Inspection of the line profiles in
Figure \ref{hratspec} also shows that the high ratio is not due to
self--absorbed \jone\ emission. We have considered optically thick
emission with temperature gradients, but some simple calculations show
that the gradients required to reproduce the observed ratio are
unrealistically large compared to those expected in photodissociation
regions. Finally, we modeled the emission produced by an ensemble of
spherical, isothermal clumps, and found that we can reproduce the
observations with fairly warm ($T_g\sim40$ K) but otherwise normal
molecular clumps ($n\sim10^4$ \percmcu, $R\sim0.1$ pc). The high
temperature required may explain why these high ratios are not more
commonly observed.

We have also used the CO data to revisit the problem of the CO--\htwo\
conversion factor in the SMC. Using a simple and robust algorithm we
found the \xco\ factor in three clouds of the complex to be
$\xco\approx 7.0 \pm 3.4 \times10^{20}$ \xcounits\ ($\xco\approx 3.5
\pm 1.7$ times the Galactic value), estimated from the CO \jtwo\ data
using virial calculations and supported by the results obtained from
the \jone\ data. This value is about $1.9$ times that which we obtain
when applying our algorithm to the Rosette Molecular Cloud and GMCs in
M~33. The result is an \xco\ that is considerably smaller than would be
expected {\it a priori} for the very metal--poor SMC, which has a
metallicity $Z\sim1/9Z_\odot$. Higher values have been obtained from a
CO/virial analysis of the complex as a whole, and higher yet from a
non-virial, far-infrared analysis of the complex. These results are
consistent with the pattern emerging from extragalactic observations
over the last decade.

\acknowledgements The research of A. D. B. and A. L. was supported by
NSF grant AST-9981308. A. D. B. would like to thank J. Simon,
R. Plambeck, and D. Hollenbach for reviewing drafts of this
manuscript. We made extensive use of the NASA/IPAC Extragalactic
Database (NED), the Los Alamos National Laboratory astrophysics
preprint database, NASA's Astrophysics Data System Abstract Service
(ADS), and the Centree de Donn\'ees astronomiques de Strasbourg (CDS)
online databases, specially the Aladin sky atlas and the Simbad
reference database.

\begin{deluxetable}{lrrr@{$\pm$}lr@{$\pm$}lr@{$\pm$}lr@{$\pm$}l}
\tablewidth{0pt}

\tablecaption{CO \jtwo\ Molecular Peaks in the N83/N84 Region \label{cloudtab}}
\tablehead{\colhead{Source}&\colhead{$\Delta \alpha$}&\colhead{$\Delta \delta$}&\multicolumn{2}{c}{T$_{peak}$}&\multicolumn{2}{c}{V$_{lsr}$}&\multicolumn{2}{c}{FWHM}&\multicolumn{2}{c}{$\int T_{mb}\,dv$}\\
&\colhead{(\arcmin)}&\colhead{(\arcmin)}&\multicolumn{2}{c}{(K)}&\multicolumn{2}{c}{(km s$^{-1}$)}&\multicolumn{2}{c}{(km s$^{-1}$)}&\multicolumn{2}{c}{(K km s$^{-1}$)}}
\tabletypesize{\footnotesize}
\startdata
MP1 & -5.6 & +5.4 & 1.08 & 0.06 & 159.29 & 0.05 & 1.96 & 0.12 & 2.2 & 0.12 \\
MP2 & -5.2 & +4.7 & 0.64 & 0.02 & 163.84 & 0.09 & 5.09 & 0.22 & 3.44 & 0.12 \\
MP3 & -4.6 & +2.2 & 0.76 & 0.04 & 158.46 & 0.08 & 2.78 & 0.19 & 2.18 & 0.12 \\
MP4 & -4.1 & +3.3 & 0.46 & 0.02 & 156.31 & 0.04 & 2.38 & 0.09 & 1.14 & 0.04 \\
MP5 & -3.9 & +4.8 & 1.84 & 0.04 & 162.26 & 0.04 & 3.34 & 0.09 & 6.4 & 0.14 \\
MP6 & -3.5 & +5.7 & 1.2 & 0.02 & 151.27 & 0.03 & 2.25 & 0.06 & 2.82 & 0.06 \\
MP7 & -2.9 & +3.9 & 2.74 & 0.02 & 162.32 & 0.01 & 2.97 & 0.03 & 8.6 & 0.06 \\
MP8 & -2.7 & +1.2 & 0.5 & 0.02 & 166.53 & 0.05 & 2.07 & 0.13 & 1.08 & 0.06 \\
MP9 & -2.4 & +5.6 & 0.22 & 0.04 & 159.48 & 0.45 & 5.61 & 1.05 & 1.22 & 0.2 \\
MP10 & -1.9 & +5.0 & 1.26 & 0.04 & 160.16 & 0.05 & 3.84 & 0.11 & 5.28 & 0.12 \\
MP11 & -0.5 & +2.2 & 0.48 & 0.02 & 167.81 & 0.11 & 4.38 & 0.27 & 2.2 & 0.12 \\
MP12 & -0.2 & +4.3 & 0.34 & 0.02 & 164.19 & 0.21 & 6.25 & 0.5 & 2.02 & 0.14 \\
MP13 & +0.0 & +0.3 & 1.42 & 0.04 & 168.01 & 0.03 & 2.71 & 0.07 & 4.06 & 0.08 \\
MP14 & +1.5 & +3.3 & 0.52 & 0.04 & 167.12 & 0.09 & 3.12 & 0.22 & 1.68 & 0.1 \\
\enddata
\end{deluxetable}

\begin{deluxetable}{rrcr@{$\pm$}lr@{$\pm$}lr@{$\pm$}lr@{$\pm$}lc}
\tablewidth{0pt}

\tablecaption{Optical Depth Data \label{spectab}}
\tablehead{\colhead{$\Delta \alpha$}&\colhead{$\Delta \delta$}&\colhead{Transition}&\multicolumn{2}{c}{T$_{peak}$}&\multicolumn{2}{c}{V$_{lsr}$}&\multicolumn{2}{c}{FWHM}&\multicolumn{2}{c}{$\int T_{mb}\,dv$}&$\tau$\\
\colhead{(\arcmin)}&\colhead{(\arcmin)}&&\multicolumn{2}{c}{(K)}&\multicolumn{2}{c}{(km s$^{-1}$)}&\multicolumn{2}{c}{(km s$^{-1}$)}&\multicolumn{2}{c}{(K km s$^{-1}$)}}
\tabletypesize{\footnotesize}
\startdata
 -5.6 & +5.6 & $^{12}$CO (2--1) & 1.34 & 0.04 & 159.26 & 0.03 & 1.8 & 0.07 & 2.52 & 0.08 &\\
&& $^{13}$CO (2--1) & 0.22 & 0.08 &\multicolumn{2}{c}{\nodata} &\multicolumn{2}{c}{\nodata} & 0.1 & 0.04 & $1.0-3.5$ \\
 +0.0 & -0.4 & $^{12}$CO (2--1) & 0.78 & 0.04 & 169.04 & 0.06 & 2.57 & 0.15 & 2.08 & 0.1 &\\
&& $^{13}$CO (2--1) & 0.04 & 0.004 & 169.02 & 0.15 & 2.48 & 0.35 & 0.1 & 0.012 & $1.5-4.4$ \\
 +0.2 & +0.6 & $^{12}$CO (2--1) & 1.66 & 0.06 & 167.98 & 0.03 & 1.74 & 0.07 & 2.96 & 0.02 &\\
&& $^{13}$CO (2--1) & 0.32 & 0.04 & 167.91 & 0.1 & 1.3 & 0.23 & 0.42 & 0.06 & $6.1-13.8$ \\
 -1.6 & +5.2 & $^{12}$CO (2--1) & 1.14 & 0.04 & 160.32 & 0.07 & 3.89 & 0.16 & 4.76 & 0.16 &\\
&& $^{13}$CO (2--1) & 0.16 & 0.02 & 160.63 & 0.17 & 4.58 & 0.41 & 0.82 & 0.06 & $7.6-17.0$ \\
 -2.8 & +3.8 & $^{12}$CO (2--1) & 3.04 & 0.02 & 162.45 & 0.02 & 3.07 & 0.03 & 9.84 & 0.14 &\\
&& $^{13}$CO (2--1) & 0.4 & 0.02 & 162.05 & 0.05 & 2.37 & 0.11 & 0.96 & 0.06 & $4.0-9.2$ \\
 -3.6 & +4.6 & $^{12}$CO (2--1) & 3.5 & 0.02 & 161.87 & 0.01 & 2.57 & 0.01 & 9.34 & 0.12 &\\
&& $^{13}$CO (2--1) & 0.4 & 0.02 & 161.81 & 0.03 & 2.36 & 0.08 & 0.98 & 0.02 & $4.4-10.0$ \\
\enddata
\end{deluxetable}

\begin{deluxetable}{lccccc}
\tablecaption{Cloud Properties \label{cotab}}
\tablehead{ \colhead{Quantity} &
\colhead{MP 7} & \colhead{MP 10} &
\colhead{MP 13} & \colhead{N. Region\tablenotemark{c}} & \colhead{Complex} }

\startdata
\sidehead{CO \jtwo}\tableline
$\alpha_{\mbox{off}}$ & $-3\farcm09$ & $-1\farcm75$ & $+0\farcm02$ &
$-3\farcm00$ & $-2\farcm34$ \\
$\delta_{\mbox{off}}$ & $+3\farcm98$ & $+5\farcm01$ & $+0\farcm39$ &
$+4\farcm35$ & $+3\farcm62$ \\
Velocity Centroid (\kmpers) & $162.0$ & $160.3$ & $168.3$ & $161.6$ & $162.0$ \\
Velocity Dispersion $\sigma_v$ (\kmpers) & 1.25 & 1.45 & 1.1 & 1.94 & 4.7/3.0\tablenotemark{a} \\
$R_{sph}$ (pc) & 22 & 14 & 13.5 & 32 & 65 \\
$S_{\rm CO}$ (Jy \kmpers) \tablenotemark{b} & 497 & 136 & 115  & 742 & 1,700 \\
M$_{\rm mol}$ from CO ($10^3$ M$_{\odot}$) & 21 & 5.8 & 4.9 & 32 & 73 \\
M$_{\rm vir}$ Virial Mass ($10^3$ M$_{\odot}$) & 35 & 29 & 18 & 126 & 1,500 \\
${\rm M_{vir}}/{\rm M_{mol}}$ & 1.7 & 5.0 & 3.7 & 4.0 & 20 \\
$\frac{\aco}{\aco(MW)}$\tablenotemark{d} & 0.9 & 2.7 & 1.9 & 2.2 & 11 \\

\tableline\sidehead{CO \jone}\tableline
$\alpha_{\mbox{off}}$ & $-3\farcm10$ & $-1\farcm73$ & $+0\farcm24$ &
$-2\farcm92$ & $-1\farcm91$ \\
$\delta_{\mbox{off}}$ & $+4\farcm02$ & $+5\farcm03$ & $+0\farcm54$ &
$+4\farcm43$ & $+3\farcm64$ \\
Velocity Centroid (\kmpers) & $162.2$ & $160.0$ & $168.4$ & $161.5$ & $162.2$ \\
Velocity Dispersion $\sigma_v$ (\kmpers) & 1.2 & 2.0 & 1.25 &
2.2/1.8\tablenotemark{a} & 4.2/3.2\tablenotemark{a} \\
$R_{sph}$ (pc) & 21 & 15 & 16
& 35 & 72 \\
$S_{\rm CO}$ (Jy \kmpers) & 440 & 193 &
132 & 715 & 1,500 \\
M$_{\rm mol}$ from CO ($10^3$ M$_{\odot}$) & 19 & 8.2 &
5.6 & 30 & 63 \\
M$_{\rm vir}$ Virial Mass ($10^3$ M$_{\odot}$) & 32 & 60 &
25 & 176 & 1,300 \\
${\rm M_{vir}}/{\rm M_{mol}}$ & 1.7 & 7.3 & 4.5 & 5.8 & 21 \\
$\frac{\aco}{\aco(MW)}$\tablenotemark{d} & 0.9 & 3.9 & 2.4 & 3.1 & 11 \\
\enddata

\tablenotetext{a}{Dispersion from the second moment/from the gaussian
equivalent width. The two differ because the line is not well
described by a single gaussian. Unless specified, the dispersion
referred to in the text is that obtained from the second moment.}

\tablenotetext{b}{The \co\ \jone\ intensity is derived from the
\jtwo\ measurements, using $\Ico\jone = 0.9\,\Ico\jtwo$.}

\tablenotetext{c}{The Northern Region is defined to include the clouds
associated with N84C and N83B/C, but excludes MP6.}

\tablenotetext{d}{Computed by dividing the previous row by the ratio
obtained using the same algorithm in Galactic and M~33 sources
(1.85).}

\end{deluxetable}

\end{document}